\documentclass[useAMS,usenatbib]{mn2e}

\usepackage{graphicx}
\usepackage{amsmath}
\usepackage{mathrsfs}
\usepackage[]{txfonts}
\usepackage{subfigure}
\usepackage{epsfig}
\usepackage{booktabs}
\usepackage{url}
\def\simless{\mathbin{\lower 3pt\hbox
{$\rlap{\raise 5pt\hbox{$\char'074$}}\mathchar"7218$}}}   
\def\simmore{\mathbin{\lower 3pt\hbox
{$\rlap{\raise 5pt\hbox{$\char'076$}}\mathchar"7218$}}}   
\newcommand{\be}{\begin{equation}}
\newcommand{\ee}{\end{equation}}
\newcommand       \bea          {\begin{eqnarray}}
\newcommand       \eea          {\end{eqnarray}}
\newcommand       \apj          {ApJ}
\newcommand       \apjl         {ApJL}
\newcommand       \aap          {A\&A}
\newcommand       \nat          {Nature}
\newcommand       \mnras        {MNRAS}
\newcommand       \pasp      {PASP}

\newcommand       \prd      {Phys.~Rev.~D.~}

\newcommand       \araa      {ARA\&A}

\newcommand \physrep {Physics Reports}
\newcommand \gca {Geochimica et Cosmochimica Acta} 

\def\simlt{\mathrel{\hbox{\rlap{\hbox{\lower4pt\hbox{$\sim$}}}\hbox{$<$}}}}
\def\simgt{\mathrel{\hbox{\rlap{\hbox{\lower4pt\hbox{$\sim$}}}\hbox{$>$}}}}
\def\lesssim{\mathrel{\hbox{\rlap{\hbox{\lower4pt\hbox{$\sim$}}}\hbox{$<$}}}}

\def\gtrsim{\mathrel{\hbox{\rlap{\hbox{\lower4pt\hbox{$\sim$}}}\hbox{$>$}}}}

\title[X-ray lines from $r$-process nuclei]{X-ray decay lines from heavy nuclei in supernova remnants as a probe of the $r$-process origin and the birth periods of magnetars}

\author[]{Justin~L.~Ripley$^{1}$, Brian~D.~Metzger$^{1}$$\thanks{E-mail: bmetzger@phys.columbia.edu}$, Almudena Arcones$^{2,3}$, Gabriel Mart{\'{\i}}nez-Pinedo$^{2,3}$\\
$^{1}$Department of Physics and Columbia Astrophysics Laboratory, Columbia University, New York, NY, 10027\\
$^{2}$ Institut f\"{u}r Kernphysik, Technische Universit\"{a}t Darmstadt, 64291 Darmstadt, Germany \\
$^{3}$ GSI Helmholtzzentrum f\"{u}r Schwerionenforschung, Planckstr.~1, 64291 Darmstadt, Germany\\}

\begin{document}
\date{Received / Accepted}
\pagerange{\pageref{firstpage}--\pageref{lastpage}} \pubyear{2012}

\maketitle

\label{firstpage}

\begin{abstract}

The origin of rapid neutron capture ($r$-process) nuclei remains one of the longest standing mysteries in nuclear astrophysics.  Core collapse supernovae (SNe) and neutron star binary mergers are likely $r$-process sites, but little evidence yet exists for their {\it in situ} formation in such environments.  Motivated by the advent of sensitive new or planned X-ray telescopes such as the Nuclear Spectroscopic Telescope Array (NuSTAR) and the Large Observatory for X-ray Timing (LOFT), we revisit the prospects for the detection of X-ray decay lines from $r$-process nuclei in young or nearby supernova remnants.  For all remnants planned to be observed by NuSTAR (and several others), we conclude that $r$-process nuclei are detectable only if the remnant possesses a large overabundance $\mathscr{O} \gtrsim 10^3$ relative to the average yield per SN.  Prospects are better for the next Galactic SN (assumed age of 3 years and distance of 10 kpc), for which an average $r$-process yield is detectable via the 10.7(9.2) keV line complexes of $^{194}$Os by LOFT at 6$\sigma$(5$\sigma$)  confidence; the 27.3 keV line complex of $^{125}$Sb is detectable by NuSTAR at 2 $\sigma$ for $\mathscr{O} \gtrsim 2$.  We also consider X-rays lines from the remnants of Galactic magnetars, motivated by the much higher $r$-process yields of the magneto-rotationally driven SNe predicted to birth magnetars.  The $\sim$ 3.6-3.9 keV lines of $^{126}$Sn are potentially detectable in the remnants of the magnetars 1E1547.0-5408 and 1E2259+586 by LOFT for an assumed $r$-process yield predicted by recent simulations.  {\it The (non-)detection of these lines can thus probe whether magnetars are indeed born with millisecond periods.}  Finally, we consider a blind survey of the Galactic plane with LOFT for $r$-process lines from the most recent binary neutron star merger remnant, concluding that a detection is unlikely without additional information on the merger location.

\end{abstract} 
  

\section{Introduction} 
\label{intro}

Nucleosynthesis via neutron capture is responsible for essentially all naturally occurring isotopes above the iron group \citep{Burbidge+57, Cameron57}.  Roughly half of these can be accounted for via the rapid ($r$) process, in which heavy nuclei are produced by capturing neutrons on a timescale shorter than their beta-decay lifetime.  Although the basic physics of the $r$-process has been understood for over half a century, its astrophysical origin remains a mystery (see \citealt{Qian&Wasserburg07}; \citealt{Arnould+07}; \citealt{Thielemann+11} for recent reviews).  In particular, direct and definitive evidence for the {\it in situ} production of $r$-process elements is currently lacking (although see \citealt{Berger+13}; \citealt{Tanvir+13}). 

The two main candidates for the site of the $r$-process are core collapse supernovae (SNe) (e.g.~\citealt{Meyer+92}; \citealt{Takahashi+94}; \citealt{Woosley+94}) and neutron star binary mergers (e.g.~\citealt{Lattimer&Schramm74}; \citealt{Freiburghaus+99}; \citealt{Korobkin+12}; \citealt{Fernandez&Metzger13}).  Evidence based on the abundances of $r$-process elements in metal-poor stars suggests that SNe contribute at least some of the $r$-process (e.g.~\citealt{Mathews+92}; \citealt{Ishimaru&Wanajo99}; \citealt{Qian00}; \citealt{Argast+04}), especially early in the chemical evolution of the Galaxy.  However, current calculations of nucleosynthesis in proto-neutron star winds fail to produce a sufficient quantity of heavy (mass number $A \gtrsim 130$) $r$-process elements to account for their solar abundances (\citealt{Thompson+01}; \citealt{Roberts+10};  \citealt{Arcones&Montes11}; \citealt{Martinez-Pinedo+13}; \citealt{Wanajo13}).  One important exception are the magneto-rotationally driven SNe that may give birth to neutron stars with especially rapid rotation and strong magnetic fields (`millisecond magnetars'), which calculations show produce a much larger quantity of $r$-process material than is possible in normal SNe (\citealt{Thompson03}; \citealt{Metzger+08}; \citealt{Ono+12}; \citealt{Winteler+12}).  Neutron star binary mergers undoubtedly contribute to the Galactic $r$-process at some level, but whether they are the dominant production channel remains unknown due to the uncertain Galactic rate of mergers (\citealt{Kim:2004hua}).

Most freshly synthesized $r$-process elements are radioactive.  As pointed out by \citet{Qian+98}, the detection of gamma-rays from decaying $r$-process nuclei in young SN remnants is a potentially promising probe of their origin (see also \citealt{Qian+99}).  Decay lines from the [non $r$-process] isotope $^{44}$Ti have been detected in SN remnants Cas A (\citealt{Vink+01}; \citealt{Renaud+06}) and 1987A (\citealt{Grebenev+12}), with the $^{44}$Ti mass estimated to be $\sim 3\times 10^{-4}M_{\odot}$.  The biggest challenge in detecting $r$-process elements in SN remnants is that their expected abundances are typically at least four orders of magnitude smaller, assuming that the $r$-process is produced in equal quantity in all core collapse events.   

In this paper we revisit the estimates of \citet{Qian+98}, but with a focus on X-ray ($\lesssim$ 80 keV) instead of gamma-ray lines.  We are motivated by the recent launch of the Nuclear Spectroscopic Telescope Array (NuSTAR; \citealt{Harrison+13}), which has unprecedented sensitivity to X-ray line emission beyond the $\sim 10$ keV high-energy cutoff achieved by previous instruments.  We are also motivated by planned future instruments, such as the Large Observatory for X-ray Timing (LOFT; \citealt{Feroci+12}), the Large Area Detector (LAD) onboard which promises an effective area up to $\sim 10$ m$^{2}$ in the $\sim 2-30$ keV spectral window.  Also unlike \citet{Qian+98}, our search includes decay lines from $A \sim 90$ light element primary process [LEPP] elements, since current models predict a sizable yield of these elements in supernova remnants, even if the $r$-process itself is not achieved (\citealt{Roberts+10};  \citealt{Arcones&Montes11}; \citealt{Martinez-Pinedo+13}).  

This paper is organized as follows.  In $\S \ref{sec:emission}$ we describe our strategy for systematically searching through the decay lines of $r$-process nuclei across the periodic table, calculating their X-ray flux, and assessing their detectability.  In $\S \ref{sec:scenarios}$ we describe three scenarios for searching for $r$-process lines: (1) individual known Galactic SN remnants, assuming an equal $r$-process yield per SN ($\S\ref{sec:SN}$); (2) SN remnants hosting magnetars, assuming a much larger $r$-process yield, as may be produced in magneto-centrifugally driven SNe ($\S\ref{sec:magnetar}$);  (3) a survey of the Galactic plane for the most recent neutron star merger remnant ($\S\ref{sec:merger}$).  In $\S \ref{sec:results}$ we present our results for the most promising X-ray lines for each scenario.  In $\S\ref{sec:discussion}$ we discuss our results and summarize our conclusions.

\section{X-ray lines from the decay of $r$-process nuclei}
\label{sec:emission}

Nuclei produced by the $r$-process decay mainly through a chain of $\beta$-decays, in some cases fed by the fission of long lived trans lead nuclei, to the (chiefly stable) $r$-process elements observed in the Galaxy today.  In our analysis we consider elements on the neutron rich side of the valley of stability as possible $r$-process parent elements.  We focus on parents that decay solely via beta decay; our search thus excludes heavier $r$-process parents with atomic number $A >209$ (\citealt{Qian+99}). 

For $r$-process isotopes to be present in an appreciable abundance in the remnant, the half-lives of the parent nuclei must be comparable or longer than the age of the system.  In young supernova remnants, the decay time must also be longer than the time it takes for the ejecta to become transparent to the X-ray line in question.  At a minimum, the ejecta must be transparent to Thomson scattering, as typically occurs after a timescale of a few years.  Even a single scattering (optical depth of unity) is not permitted, since the energy shift due to the in-elasticity of the scattering process, $\Delta \epsilon_{\rm X}/\epsilon_{\rm X} \sim \epsilon_{\rm X}/m_e c^{2} \gtrsim 0.03$, generally exceeds the intrinsic line width due to the Doppler shift of the expanding ejecta and the energy resolution of the X-ray detector.  For soft X-rays $\epsilon_{\rm X} \lesssim 10$ keV, bound-free absorption by e.g., iron, also suppresses the escape of X-rays for up to a decade or longer.  Young supernovae remnants are also strong sources of soft X-ray synchrotron emission, representing a challenging background to overcome ($\S\ref{sec:flux}$).  For these reasons we focus on isotopes with decay X-ray of $\epsilon_{\rm X} \gtrsim 10$ keV when considering especially young remnants.

Table \ref{table:parents} provides the sample of $r$-process parent elements used in our analysis.  These were selected from the LBNL Isotope database\footnote{\url{http://ie.lbl.gov/toi.html}} for $r$-process nuclei with half-lives of greater than three years, X-ray lines with energies $\epsilon_X < 80$ keV, and of atomic mass $A < 209$.  We note that the decay times $\tau$ for some isotopes, particularly those with especially long decay times, are uncertain up to a factor of a few, which results in a similar uncertainty in our calculations.  

\begin{table}
\begin{scriptsize}
\begin{center}
\vspace{0.05 in}\caption{$r$-process isotopes and their parent nuclei in our sample}
\label{table:parents}
\begin{tabular}{cccccc}
\hline \hline 
\multicolumn{1}{c}{Daughter} &
\multicolumn{1}{c}{$X_{\odot} {^{(a)}}$} &
\multicolumn{1}{c}{$\chi^{(b)}$} &
\multicolumn{1}{c}{Parent} &
\multicolumn{1}{c}{$\tau^{(c)}$} \\
\hline 
\vspace{-0.3cm}
\\
 &  & &  & (yr)  \\
\hline
\\
$^{99}$Ru & $5.94 \times 10^{-10}$ & 0.169 & $^{99}$Tc & $3.04 \times 10^5$ & \\
$^{125}$Te & $1.06 \times 10^{-9}$ & 0.274 & $^{125}$Sb  & 3.98  \\  
$^{126}$Te & $2.87 \times 10^{-9}$ & 0.541 & $^{126}$Sn & $3.32 \times 10^5$ \\ 
$^{129}$Xe & $4.21 \times 10^{-9}$ & 0.124 & $^{129}$I & $2.27 \times 10^7$ \\
$^{137}$Ba & $1.76 \times 10^{-9}$ & 0.174 & $^{137}$Cs & $43.4$\\
$^{151}$Eu & $1.78 \times 10^{-10}$ & 0.044 & $^{151}$Sm & $1.39 \times 10^2$ \\
$^{155}$Gd & $1.91 \times 10^{-10}$  & 0.046 & $^{155}$Eu & $6.86$\\
$^{182}$W & $1.62 \times 10^{-10}$  & 0.019 & $^{182}$Hf & $1.29 \times 10^7$ \\
$^{194}$Pt & $2.16 \times 10^{-9}$  & 0.423 & $^{194}$Os & $8.66$\\
  \\
\hline
\hline
\end{tabular}
\end{center}
$^{(a)}$ Solar mass fraction of daughter isotope (\citealt{Anders&Grevesse89}). $^{(b)}$ Fraction of daughter isotope produced via $r$-process \citep{Arlandini+99}. $^{(c)}$ Lifetime of parent isotope. 
\end{scriptsize}
\end{table}

\subsection{Ejecta Mass per Supernova or Merger Event}

Assuming that either core collapse SNe or neutron star binary mergers are the dominant source of $r$-process nuclei in the Galaxy, then the average mass of a given element produced per such event can be estimated as (\citealt{Qian+98}; \citealt{Qian+99})
\begin{equation}
\label{equation:mass}
	\langle M^{r} \rangle \approx \frac{X_{\odot} M_G \chi}{\mathcal{R}\tau_G}
\end{equation}
where $X_{\odot}$ is the solar mass fraction of the daughter element, $M_G \approx 10^{11}M_{\odot}$ and $\tau_G \approx 10^{10}$ years are respectively the stellar mass and age of the galaxy, $\mathcal{R}$ is the Galactic event rate (SNe or merger), and $\chi$ is the fraction of the daughter element that results from the $r$-process (as opposed to other nucleosynthetic channels, e.g. the $s$-process).  Equation (\ref{equation:mass}) is a good approximation provided that the lifetime of the isotope in question is much shorter than the age of the solar system and of the Galaxy, as is always well satisfied.  

If the quantity of the nuclei produced in a SN or merger can be constrained by observations to be less than $\langle M^{r} \rangle$, then this begins to place interesting constraints on the production of such elements in the site of interest.  Equation (\ref{equation:mass}) thus represents our `figure of merit' when assessing the detectability of a given X-ray line signal.   


\subsection{X-ray Flux from a SN or Merger Remnant}
\label{sec:flux}

For each X-ray line of energy $\epsilon_{X}$ produced by the decay of the $r$-process element $r$, the fiducial value of the X-ray line flux from the SN or merger remnant is given by
\begin{equation}
\label{equation:flux}
	\langle\dot{N}_{X}\rangle = f_X \frac{\langle M^r \rangle}{AM_u}\frac{e^{-t/\tau}}{4 \pi d^2 \tau}
\end{equation}
where $f_X$ is the fraction of decays that release a photon at energy $\epsilon_{\rm X}$, $A M_u$ is the mass of the parent nucleus, $\tau$ is the lifetime of the parent nucleus, and $d$ is the distance to the supernova/neutron star merger, and $\langle M^r \rangle$ is the total mass of the $r$-process isotope (eq.~[\ref{equation:mass}]).  Everything else being equal, the flux $\langle\dot{N}_{X}\rangle$ is maximal for an isotope with a decay time equal to the age of the system ($\tau = t$).  Note that we do not expect the X-ray line to be resolved, because the energy resolution of NuSTAR or LOFT exceeds the expected line width $\Delta \epsilon_{X}/\epsilon_{X} \sim v/c \lesssim 0.02$ as set by the Doppler broadening to the outwardly expanding ejecta, where $v \lesssim 6,000$ km s$^{-1}$ is the velocity of the supernova or merger ejecta at the time of observation\footnote{Although the velocity of the ejecta from a neutron star binary merger is initially significantly higher ($v \gtrsim 0.1 c$), by now the ejecta will have slowed to $v \ll$ few 10$^{3}$ km s$^{-1}$.}. 

The `Galactic average' line flux in equation (\ref{equation:flux}) is insufficient for a detection in most scenarios that we consider.  We nevertheless scale the actual line flux $\dot{N}_X$ to this value according to
\be
	\dot{N}_X = \mathscr{O}\langle\dot{N}_{X}\rangle,
\label{equation:fluxO}
\ee
where $\mathscr{O} \equiv M^{r}/\langle M^r \rangle$ is the `overabundance', which we define as the ratio of the actual mass of the $r$-process element synthesized $M^{r}$ relative to its average yield per SN or merger event $\langle M^r \rangle$ (eq.~[\ref{equation:mass}]).  Hence $\mathscr{O}$ can be $>1$ or $< 1$, depending on whether the isotope in question is over- or under-abundant relative to the averaged expected yield per SN or merger event.

\subsection{Quantifying Detection Confidence}
\label{confidence}

We approximate the ratio of signal to noise required for detection of an X-ray line by the statistical method known as student's t-test.  The t-statistic for detecting the line flux (signal) over the random fluctuation of the background (noise) is estimated as follows (see Appendix \ref{sec:append} for a detailed derivation)
\begin{equation} 
\label{equation:ttestexp}
	t_{X}(\mathscr{O}) = \frac{\dot{N}_{X} \sqrt{A_{X}T}}{\sqrt{\dot{N}_{X} + 2 \nu_X}}
\end{equation}
where $\dot{N}_{X}$ is the signal flux, $\nu_X$ is the rate of detector noise, $A_{X}$ is the effective area of the detector at energy $X$, and $T$ is the observation duration. The instrumental noise $\nu_X$ and effective areas for NuSTAR and LOFT were estimated from \citet{Harrison+13} (their Table 2 and Figs.~2 and 10) and \citet{Belloni:2012jc} (their Table 1 and Figure 3), respectively.  When lines of the same isotope have energies that overlap within the energy resolution of the instrument, they are combined into a single flux for purposes of calculating their signal.  The energy resolution for NuSTAR and LOFT were obtained from Table 2 of \citet{Harrison+13} and Table 2 of \citet{Belloni:2012jc}, respectively.  Our calculation of the flux and background also accounts for the relative angular size $\theta^2$ of the $r$-process source relative to the field of view (FoV) $\Theta^2$ of NuSTAR or LOFT.  If the FoV is smaller than the angular size of NuSTAR/LOFT, then the total flux $\dot{N}_X$ is suppressed by a factor $\theta^2 /\Theta^2$; we thus assume that $r$-process flux is uniform across the solid angle of the remnant.  Likewise, if the angular size of the source was smaller than the FoV, then the background counts $\nu_X$ is suppressed by $\Theta^2/\theta^2$ relative to its total across the entire FoV.

The t-statistic $t_X$ for each line corresponds to the probability $\mathscr{C}_X$ (`confidence value') that the measured photon flux over the observation is due entirely to random background fluctuations.  The complementary probability $\mathscr{P}_X = 1 - \mathscr{C}_X$ corresponds to the observed X-ray flux not being due to random noise, i.e.~a detection.  In most cases we express our results in terms of the overabundance $\mathscr{O}$ (eq.~[\ref{equation:fluxO}]) required for each line in order to to achieve a chosen value $\mathscr{P}_X \sim 0.95$, corresponding to a moderately high ($2\sigma$) detection confidence.  

Our analysis only takes into account detector noise $\nu_X$, but does not account for other astrophysical backgrounds such as the X-ray continuum of the SN remnant itself.  Essentially all Galactic SNRs less than 2000 years old produce synchrotron X-ray emission (\citealt{Reynolds08}).  However, usually the synchrotron spectrum reaches a peak at high energy $\epsilon_{\rm X} \sim $ few keV and decreases as a power law or steeper at higher energies (e.g.~\citealt{Patnaude&Fesen09}; \citealt{Matheson&SafiHarb10}; \citealt{Sturm+10}).  In SN 1987A,  for instance, the continuum X-ray flux at energies $\geq 10$ keV under consideration is $\lesssim 10^{-4}$cts cm$^{-1}$ s$^{-1}$ keV$^{-1}$, which is smaller than the instrument noise for NuSTAR and LOFT.  Thus, given the caveat that continuum emission may provide an additional obstacle to detection in some systems, equation (\ref{equation:ttestexp}) nevertheless provides a rough optimistic check on the feasibility of detecting a single X-ray line using a telescope with characteristic noise $\nu_X$.  In the case of magnetar remnants, one must also be able filter out X-rays from the magnetar itself, as is usually possible if the angular resolution of the detector is smaller than the size of the remnant or if the observations are taken when the magnetar is in quiescence.

In some cases we also consider the enhanced sensitivity to detecting $r$-process lines that may be achieved by combining the statistical significance of signals from multiple separate lines.  This approach is motivated by the fact that $r$-process isotopes are unlikely to be produced in isolation due to the nature of their formation as the result of mutiple subsequent neutron captures.  The statistics of multiple lines is not explicitly taken into account by equation (\ref{equation:ttestexp}), but the details of our methodology for this case is described in Appendix \ref{sec:append}.

\section{Search Scenarios}
\label{sec:scenarios}

Three search scenarios are considered, based on the two possible astrophysical sites of the $r$-process: supernovae (SNe) and binary neutron star mergers (NS mergers). The first scenario is a search of individual normal SN remnants ($\S\ref{sec:SN}$), chosen based on their youth (e.g.~1987 A), proximity (e.g.~Vela Jr.), and/or planned observation by NuSTAR.  The second scenario ($\S\ref{sec:magnetar}$) is a search of Galactic magnetar SN remnants, motivated by the potentially larger $r$-process yields of the magneto-rotationally driven SNe thought to birth magnetars (\citealt{Winteler+12}).  The third scenario is a blind search for the line signal from the most recent Galactic binary NS merger remnant ($\S\ref{sec:merger}$).  Mergers are expected to produce a larger quantity of $r$-process elements than SNe per event, but since their rate is lower, the most recent remnant is likely older.  Furthermore, since the location of the remnant is unknown {\it a priori}, such a signal can only be detected via a systematic survey of the Galactic plane.  The sources in our sample are summarized in Table \ref{table:sources}.

\begin{table}
\def\arraystretch{1.5}
\begin{scriptsize}
\begin{center}
\vspace{0.05 in}\caption{Summary of the $r$-process sources considered}
\label{table:sources}
\begin{tabular}{cccccc}
\hline \hline 
\multicolumn{1}{c}{Name} &
\multicolumn{1}{c}{Age} &
\multicolumn{1}{c}{Distance} &
\multicolumn{1}{c}{Angular Size} &
\multicolumn{1}{c}{Reference} \\
\hline 
\vspace{-0.3cm}
\\
 &  (yr)& (kpc) & (arcmin)  \\
\hline
\\
Vela Jr & $750^{+8300}_{-50}$ & $0.2^{+2.7}_{-0.1}$ & 120 & 1, 2\\
Cas A$\dagger$ & $320^{+32}_{-4}$ & $3.4^{+0.2}_{-0.1}$ & 5 & 1,2\\
SN 1987A$\dagger$ & 26 & $51.0^{+1.8}_{-0.6}$ & 0.5 & 3\\
SN 1006$\dagger$ & 1007 & $1.7^{+0.5}_{-0.1}$ & 30 & 1,2 \\
G001.9+00.3$\dagger$ & $160^{+60}_{-10}$ & $8.6^{+0.7}_{-0.1}$ & 1.5 & 1,2 \\
3yrs 10kpc & 3 & 10 & 0.5 & - \\
1E 2259+586$\dagger$ & $8000^{1700}_{-100}$ & $3.2^{+0.2}_{-0.2}$ & 1.9 & 4,5 \\
SGR 0501+4516 & $6200^{+1800}_{-200}$ & $0.9^{+0.3}_{-0.1}$ & 120 & 6,7 \\
1E 1841-045 & $1300^{+200}_{-200}$ & $7$ & 2.1 & 4,5\\
1E 1547.0-5408 & $1400^{+400}_{-50}$ & $3.8^{+0.7}_{-0.1}$ & 5 & 8,9 \\
AX J1845-0258 & $5000^{+3000}_{-3000}$ & $10$ & 1.1 & 5,10 \\
CXOU J171405.7–381031 & $1100^{+900}_{-100}$ & $7^{+6.7}_{-0.3}$ & 6 & 11, 12\\ 
4U 0142+61 & $1000^{+500}_{-50}$ & $3.7^{+0.3}_{-0.1}$ & 20 & 13 \\
SGR 0526-66 & $6300^{+1000}_{-1000}$ & $50$ & 0.1 & 4,5 \\
NS Merger & $10^4$ & 10 & 60 & - \\
\hline
\hline
\end{tabular}
\end{center}
$^{1}$\citet{Ferrand:2012jh}: \url{http://www.physics.umanitoba.ca/snr/SNRcat }. 
$^{2}$\citet{Green09arevised}: \url{http://www.mrao.cam.ac.uk/surveys/snrs/}.
$^{3}$\citet{Panagia:2003rt}. 
$^{4}$\citet{Vink2008503}. 
$^{5}$\citet{Woods:2004kb}. 
$^{6}$\citet{2007A&A...461.1013L}. 
$^{7}$\citet{refId0}.
$^{8}$\citet{Gelfand:2007tz}. 
$^{9}$\citet{Camilo:2007jn}
$^{10}$\citet{1999ApJ...526L..37G}
$^{11}$\citet{1674-4527-11-6-001}. 
$^{12}$\citet{Halpern:2010ce}.  
$^{13}$\citet{0004-637X-772-1-31}. 

Uncertainties ($\pm$) indicate the full range from the cited references.  Age values for most remnants are calculated with respect to the year 2013, as would apply to observations by NuSTAR.  Distances and ages adopted in our baseline calculations are taken to be the lowest from the allowed range.  $\dagger$Currently planned to be observed by NuSTAR (\citealt{Harrison+13}).  
\end{scriptsize}
\end{table}

For each scenario, we calculate the t-statistic $t_{X}$ (equation \ref{equation:ttestexp}) for the X-ray lines of each r-process element in our sample (Table \ref{table:parents}) and the resulting probability $\mathscr{P}_X$ of distinguishing the signal flux from noise.  As most of the detection probabilities $\mathscr{P}_{X}$ were very small ($\ll 1$) for the fiducial average flux (eq.~[\ref{equation:flux}]), we then inverted $t_{X}(\mathscr{O})$ to determine the overabundance required to achieve a detection probability $\mathscr{P}_{X}\sim$95\% confidence ($t_X = 1.9$).  An overabundance $\mathscr{O} < 1$ indicates that the astrophysical source could underproduce that element relative to the Galactic average and LOFT/NuSTAR could still detect a signal at $\mathscr{P}_X = 95\%$ confidence.  For SN remnants, uncertainties in the required value of $\mathscr{O}$ are dominated by the uncertainties in the remnant age $t$ and the distance $d$.  When the age and distance are well known (or assumed), uncertainties in $\mathscr{O}$ are instead dominated by uncertainties in the $r$-process fraction $\chi$ (\citealt{Arlandini+99}). 

\subsection{Normal Supernova Remnants}
\label{sec:SN}

The photon flux for the X-ray lines from the $r$-process parent elements listed in Table \ref{table:parents} were calculated using equation (\ref{equation:flux}) for several of the most promising Galactic SN remnants as well as from SN 1987A.  We assume a Galactic SN rate $\mathcal{R}= \mathcal{R}_{\rm SN}$ = $0.032^{+0.073}_{-0.026}$ yr$^{-1}$ (\citealt{Adams:2013ana}).  The SNe were  selected (Table \ref{table:sources}) primarily for being young, with ages generally less than the longest lifetime of the parent elements in Table \ref{table:parents}.  All remnants currently planned to be observed by NuSTAR were included (Table 5 of \citealt{Harrison+13}). 

We also considered the possibility that a Galactic supernova will occur and be detected in the next several years, when either NuSTAR (this decade) or LOFT (next decade) is functional.  We consider the detected X-ray signal at $t = 3$ years, since this is approximately the timescale required for the ejecta to become transparent to X-rays.  We place the hypothetical future SN at a typical distance of 10 kpc.  An observation time $T = 10^6$ s is assumed for each remnant.

\subsection{Magnetar Supernova Remnants}
\label{sec:magnetar}

It has long been speculated that magnetars achieve their strong magnetic fields as the result of being formed with extremely short $\sim$ millisecond rotation periods (\citealt{Thompson&Duncan93}), although the energetics of magnetar SN remnants have challenged this model (\citealt{Vink&Kuiper06}).  We consider another test of the millisecond birth periods of magnetars based on the predicted high $r$-process yields of their magneto-centrifugally driven SNe.

We calculate the fiducial $r$-process abundances $\langle M^r \rangle$ in the magnetar scenario assuming a {\it total} $r$-process yield of $M^r \approx 6\times 10^{-3}M_{\odot}$, as found by recent simulations of magneto-rotational SNe (\citealt{Winteler+12}); this is equivalent to using equation (\ref{equation:mass}) for an assumed event rate $\mathcal{R}\sim 10^{-4}$ yr$^{-1}$.  Our magnetar candidates are drawn from the online database compiled by the McGill Pulsar Group (Table \ref{table:sources}), supplemented by the magnetar candidates to be observed by NuSTAR \citep{Harrison+13}.  If no estimates of the magnetar age were available, then the spin down time $\tau_c = P/2\dot{P}$ was used as a rough age estimate; our results are relatively insensitive to this assumption since the most promising nuclei have decay times $\tau \gg \tau_c$.   An observation time $T = 10^6$ s is assumed for each remnant.

\subsection{Neutron Star Merger Remnant}
\label{sec:merger}

Neutron star binary mergers (NSMs) occur in the Milky Way with an estimated frequency of $\mathcal{R}_{\rm NSM} \sim 10^{-4}$ yr$^{-1}$ (\citealt{Kim:2004hua}) that is approximately $\sim 300$ times lower than that of core collapse SNe.   If NS mergers are the dominant Galactic $r$-process source, then the $r$-process mass ejected per NSM must be $\sim 300$ times higher, as follows from equation (\ref{equation:mass}) by replacing $ \mathcal{R}_{\rm SN}$ with $ \mathcal{R}_{\rm NSM}$.  Although this higher ejecta mass provides a potentially larger X-ray signal, this benefit is mitigated by our ignorance of the location of the most recent Galactic NSM and the relatively old age of the youngest remnant, which is typically $t_{\rm NSM} = \mathcal{R}_{\rm NSM}^{-1} \sim 10^4$ years.

The remnant produced by the most recent NSM should still be well-localized within the Galaxy, provided that the NS merger itself occurred in the Galactic plane.  Mass ejected by the merger begins to decelerate once it sweeps up its own mass worth of interstellar matter.  After this time (typically less than a year), the evolution of the ejecta radius approaches that of a Sedov-Taylor blastwave (\citealt{1959sdmm.book.....S})
\begin{equation}
\label{equation:sedov}
r_{\rm NSM}(t) \sim \left( \frac{Et^2}{\rho} \right)^{1/5}
\end{equation}
where $E$ is the energy of the ejecta and $\rho$ is the density of the ISM.  For typical values $ E \sim 10^{51}$ ergs (\citealt{2013PhRvD..87b4001H}; \citealt{Fernandez&Metzger13}) and $\rho \sim 10^{-24}$ g cm$^{-3}$, one finds $r_{\rm NSM}(t_{\rm NSM}) \sim 20$ pc, which is much less than the vertical thickness of the Galaxy.  

Since the location of the remnant is unknown, detecting its X-ray decay signal requires a systematic search of the Galactic plane for spectral lines of parent nuclei listed with half-lives $\sim t_{\rm NSM}$ (Table \ref{table:parents}).  Performing such a search over a reasonable time-frame requires an X-ray detector with a large FoV.  Such a search is feasible in principle with the Large Area Detector on LOFT, because its FoV $\sim 1$ deg$^{2}$ (\citealt{Belloni:2012jc} Table 1) is larger than the thickness of the Galactic plane at a typical distance $\sim 10$ kpc.  

The second scenario we consider is thus a Galactic plane survey by LOFT for the most recent NSM (remnant age $t = 10^{4}$ years; model NSM).  We assume a $10^{6}$ s integration per pointing, allowing a complete search over the course of a few years.  

\section{Results}
\label{sec:results}

Figures \ref{figure:LOFTVelaJr}$-$\ref{figure:LOFTNSM} shows the overabundance $\mathscr{O}_X = M^{r}/ \langle M^{r}  \rangle$ required for a $\mathscr{P}\geq$95\% (2 $\sigma$) confidence detection with NuSTAR or LOFT for the most promising X-ray lines in a sample of the sources that we consider.  In Figures \ref{figure:NuSTAR3yrs10kpc} and \ref{figure:LOFT1E2259}, the bottom two sub-panels also show the signal flux $\langle \dot{N}_X \rangle$ and signal-to-noise ratio $\langle \dot{N}_X \rangle / \nu_X$ for the fiducial abundance case ($\mathscr{O}_X = 1$).  Tables \ref{table:totalLOFT} and \ref{table:totalNuSTAR} summarizes our results for the most promising lines in each source for LOFT and NuSTAR, respectively.  
In addition to the required overabundance to detect individual lines $\mathscr{O}_X$, we also present the overabundance required for detection by combining the expected signal of all $r$-process lines in the final column.  In some cases we find this can enhance the signal (decrease the required overabundance) by almost an order of magnitude as compared to the most promising individual line.


\subsection{Normal Supernova Remnants}

\begin{figure}
\centering
\includegraphics[width=0.5\textwidth]{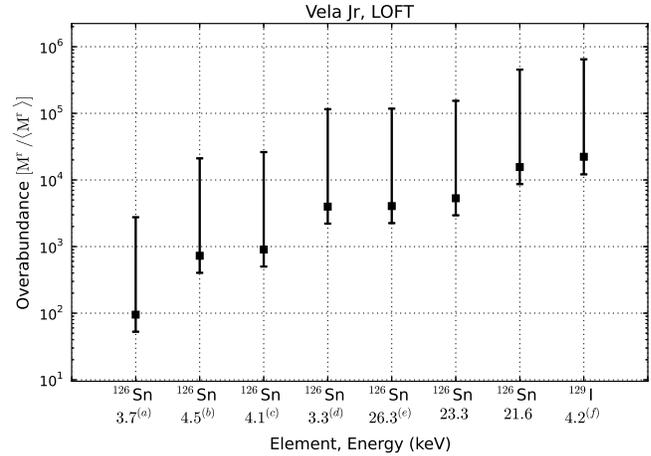}
\caption{Required overabundance $\mathscr{O} = M^{r}/ \langle M^{r}  \rangle$ for a $\mathscr{P}\geq$95\% (2 $\sigma$) confidence detection of $r$-process lines from the Vela Jr. remnant with LOFT as a function of the parent nucleus.  X-ray lines with superscripts indicate line complexes within the energy resolution of the detector which have been combined into a single line for purposes of detectability, with the individual line energies (in keV) given as follows: $^{(a)}$3.59,3.61,3.84,3.89. $^{(b)}$4.35,4.60. $^{(c)}$3.93,3.98,4.10. $^{(d)}$3.19,3.44,3.60. $^{(e)}$26.11,26.36. $^{(f)}$3.95,4.11,.}
\label{figure:LOFTVelaJr}
\end{figure}

Figure \ref{figure:LOFTVelaJr} shows our results for the Vela Jr remnant (RX J0852.0- 4622) as observed by LOFT.\footnote{The angular size of Vela Jr. ($\theta^{2} \sim$ few deg$^{2}$) is significantly larger than the FOV of NuSTAR ($\Theta^{2} \sim 0.04$ deg$^{2}$), suppressing the signal by a large factor $\Theta^{2}/\theta^2 \sim 10^{2}$ ($\S$ \ref{confidence}).  However, the angular size is coincidentally similar to the FoV of LOFT.}  The most promising lines for LOFT are from $^{126}$Sn ($\tau = 3.3\times 10^{5}$ years) at $\epsilon_X = $ 3.7, 4.5 and 4.1  keV; these are detectable for overabundances $\mathscr{O}_X \gtrsim 800$, $5 \times 10^2$, and $6 \times 10^2$, respectively.  These numbers are pessimistic despite the close proximity and relatively young age of Vela Jr. because the effective area of LOFT is sharply peaked at X-ray energies $\lesssim10$ keV around which there are no lines that contribute a sizable flux from isotopes with a half-life comparable to the age of the remnant $\sim$ 10$^{3}$ years. 

\begin{figure*}
\centering
\includegraphics[width=1\textwidth]{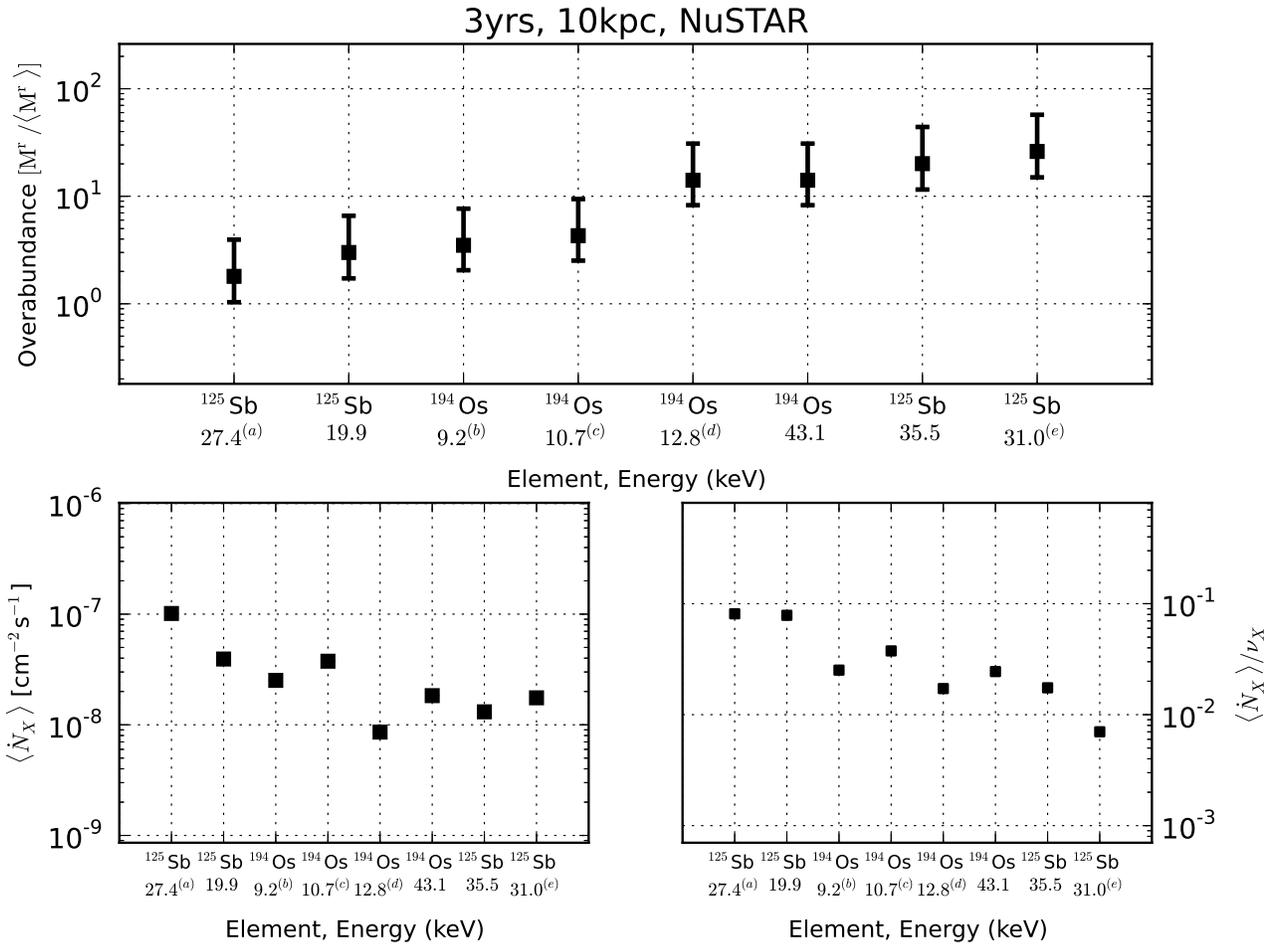}
\caption{Same as Figure \ref{figure:LOFTVelaJr}, but for a future 3 year old Galactic SNe at a distance of 10 kpc observed by NuSTAR. The bottom sub-panels show the signal flux $\langle \dot{N}_X \rangle$ and signal-to-noise ratio $\langle \dot{N}_X \rangle / \nu_X$ for the fiducial abundance case ($\mathscr{O} = 1$).  Unresolved lines have individual energies (in keV) as follows: $^{(a)}$26.88,27.20,27.40. $^{(b)}$9.10,9.17. $^{(c)}$10.51,10.53,10.71,10.87,10.91. $^{(d)}$12.51,12.82,12.84,12.92. $^{(e)}$30.94,30.99,31.24}
\label{figure:NuSTAR3yrs10kpc}
\end{figure*}
\begin{figure}
\centering
\includegraphics[width=0.5\textwidth]{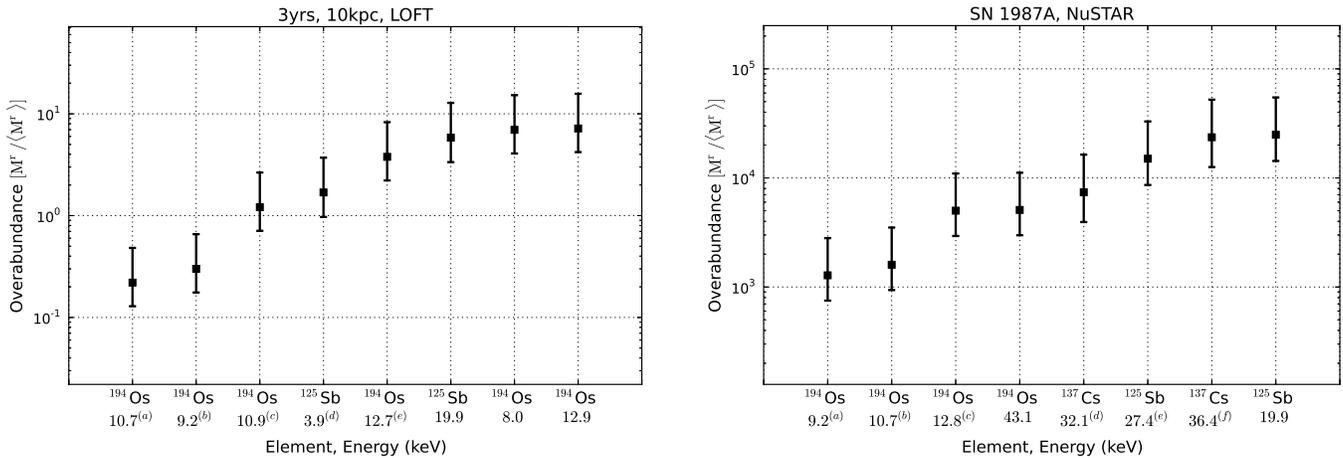}
\caption{Same as Figure \ref{figure:LOFTVelaJr}, but for a future 3 year old Galactic SNe at a distance of 10 kpc observed by LOFT. Unresolved lines have individual energies (in keV) as follows:$^{(a)}$10.51,10.53,10.71. $^{(b)}$9.10,9.17. $^{(c)}$10.87,10.91. $^{(d)}$3.76,3.77,4.03,4.07. $^{(e)}$12.51,12.82,12.84}
\label{figure:LOFT3yrs10kpc}
\end{figure}

Figures \ref{figure:NuSTAR3yrs10kpc} and \ref{figure:LOFT3yrs10kpc} shows the required overabundance for detecting $r$-process lines from a hypothetical young ($t = 3$ yrs) Galactic SN with NuSTAR and LOFT, respectively.  The most promising isotope for NuSTAR is $^{125}$Sb ($\tau = 3.98$ years), which produces several lines between 27 and 30 keV with a relatively high production probability ($f_X \sim 0.02-0.3$).  These lines are the most promising because the effective area of NuSTAR is relatively high in this band and the noise rate $\nu_X$ is relatively low.  The daughter isotope $^{125}$Te is also a relatively abundance $r$-process element.  Despite these ideal conditions, an overabundance $\mathscr \gtrsim 2$ is required for a confident detection with NuSTAR.

For LOFT the most promising isotope from the young Galactic remnant is $^{194}$Os ($\tau = 8.7$ years), which produces several lines between 9 and 12 keV with $f_X \gtrsim 0.02$.  These lines are favored because their energies overlap with the maximal effective area of LOFT and because the daughter $^{194}$Pt is relatively abundant.  Unlike NuSTAR, LOFT has a high probability (6 $\sigma$)[5$\sigma$] of detecting $^{194}$Os from the decay line complexes at 9.2 and 10.7 keV from the next Galactic SN remnant, even for the fiducial yield $\mathscr{O} = 1$.  This can be inferred from Figure \ref{figure:LOFT3yrs10kpc}, which shows that $\mathscr{O} = 0.2$ is required for a 2 $\sigma$ detection, i.e. even an {\it underabundance} by a factor of several as compared to the average yield per SN would produce detectable line emission. 

\begin{figure}
\centering
\includegraphics[width=0.5\textwidth]{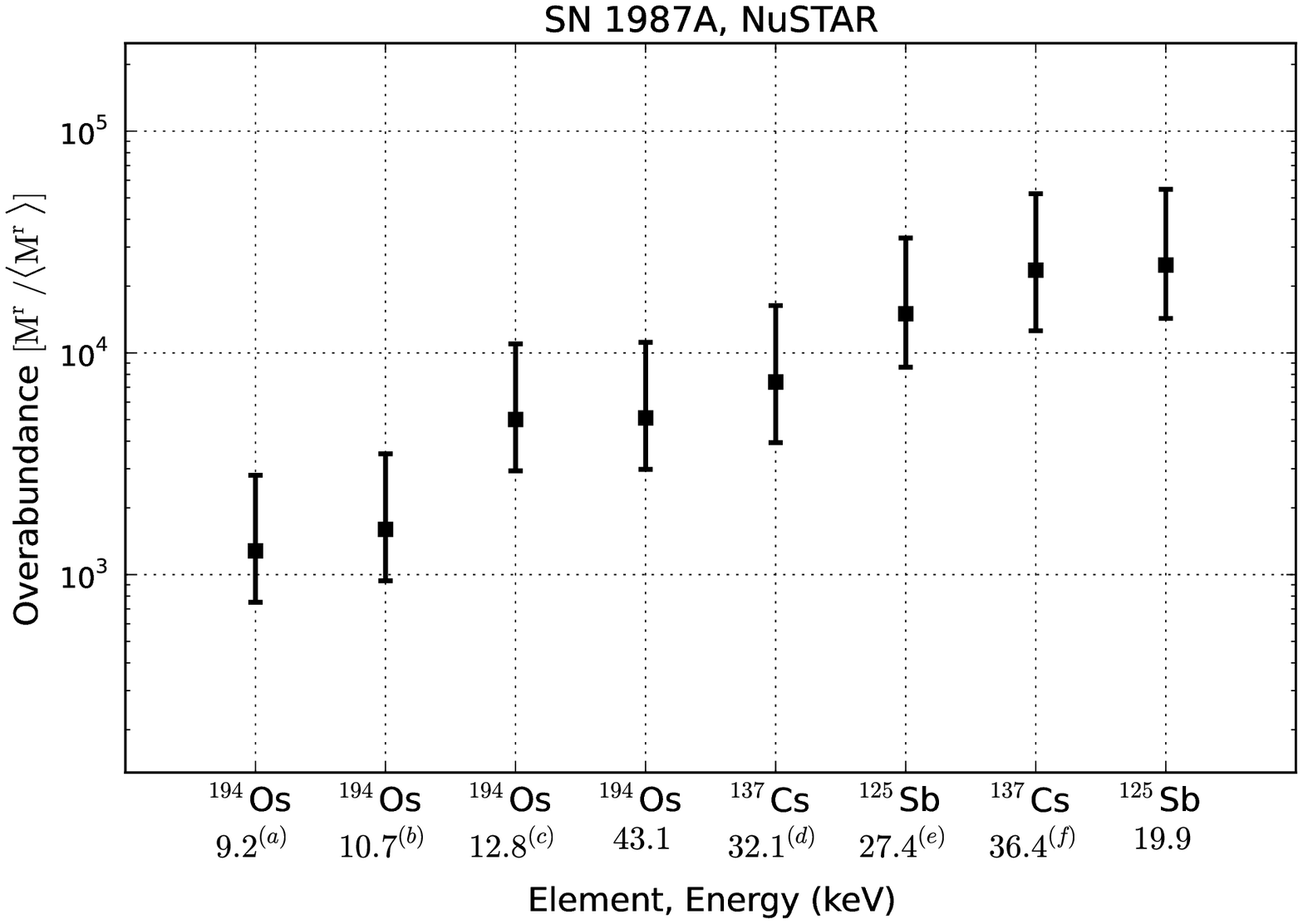}
\caption{Same as Figure \ref{figure:LOFTVelaJr}, but for SN1987A observed by NuSTAR. Unresolved lines have individual energies (in keV) as follows: $^{(a)}$9.10,9.17. $^{(b)}$10.51,10.53,10.71,10.87,10.91. $^{(c)}$12.51,12.82,12.87,12.92. $^{(d)}$31.19,32.82. $^{(e)}$27.20,27.47.$^{(f)}$36.30,36.38,36.65}
\label{figure:NuSTARSN1987A}
\end{figure}
\begin{figure*}
\centering
\includegraphics[width=1\textwidth]{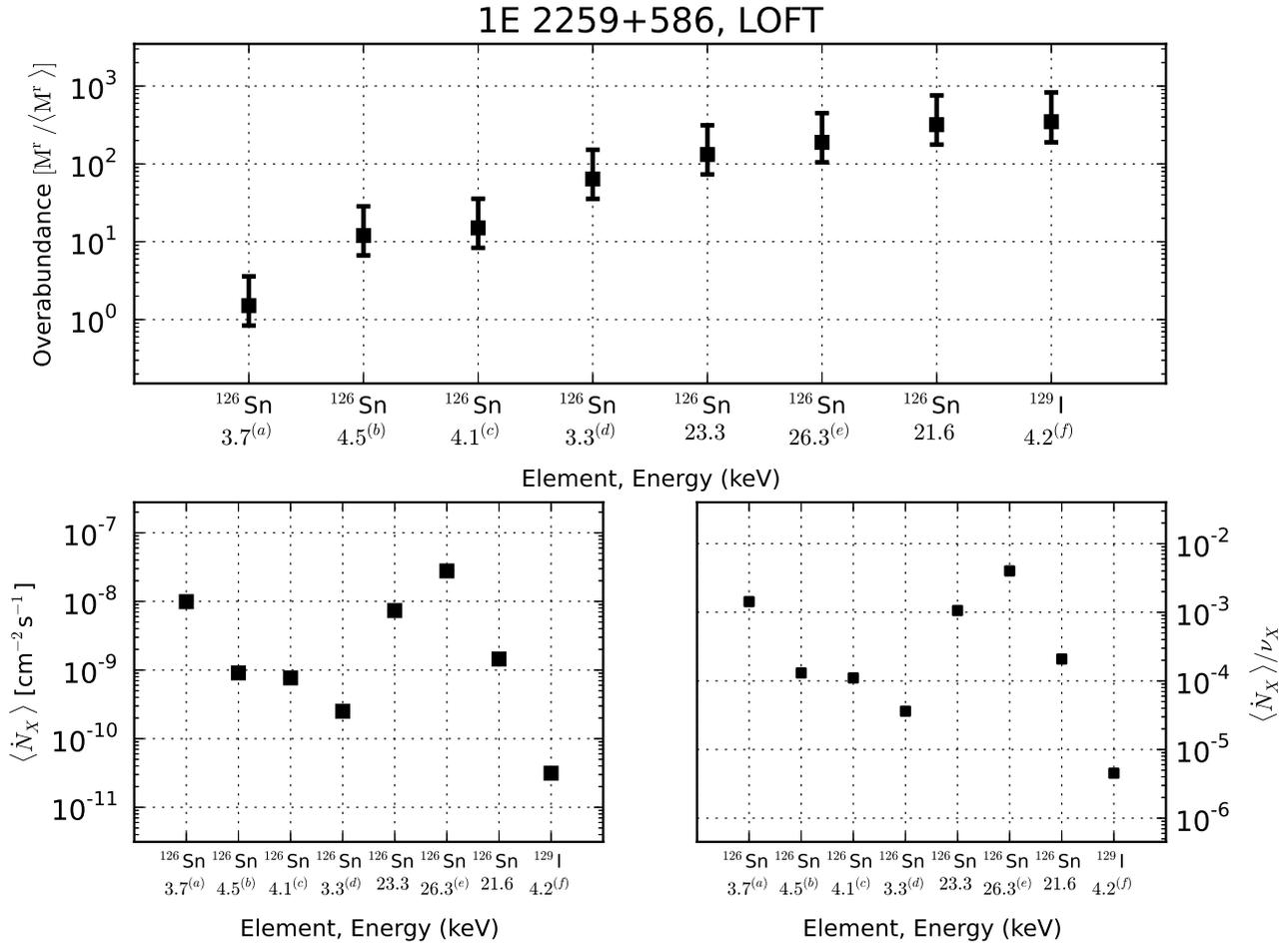}
\caption{Same as Figure \ref{figure:LOFTVelaJr}, but for a the magnetar 1E 2259+586 as observed by LOFT. The bottom subpanels show the signal flux $\langle \dot{N}_X \rangle$ and signal-to-noise ratio $\langle \dot{N}_X \rangle / \nu_X$ for the fiducial abundance case ($\mathscr{O} = 1$).  Unresolved lines have individual energies (in keV) as follows: $^{(a)}$3.61,3.84,3.89. $^{(b)}$4.35,4.60. $^{(c)}$3.93,3.98,4.10. $^{(d)}$3.19,3.44. $^{(e)}$26.11,26.36. $^{(f)}$4.09,4.11,4.40.}
\label{figure:LOFT1E2259}
\end{figure*}
\begin{figure}
\centering
\includegraphics[width=0.5\textwidth]{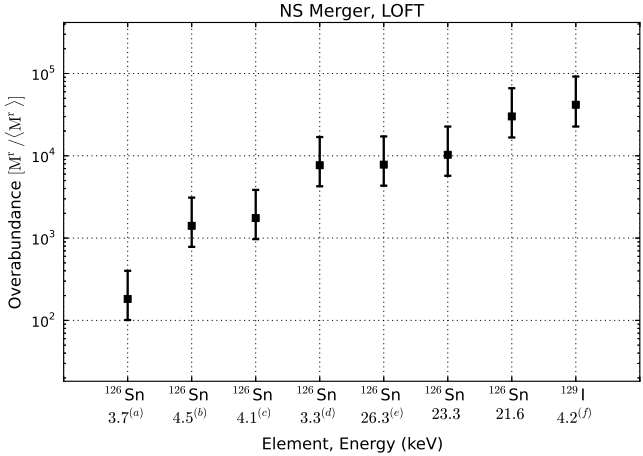}
\caption{Same as Figure \ref{figure:LOFTVelaJr}, but for a $10^{4}$ year old Galactic neutron star merger remnant at a distance of 10 kpc as observed by LOFT. Unresolved lines have individual energies (in keV) as follows: $^{(a)}$3.59,3.61,3.84,3.89. $^{(b)}$4.35,4.60. $^{(c)}$3.93,3.98,4.10. $^{(d)}$3.19,3.44,3.60. $^{(e)}$26.11,26.36. $^{(f)}$4.09,4.11,4.40.}
\label{figure:LOFTNSM}
\end{figure}

Figure \ref{figure:NuSTARSN1987A} shows the same results for an observation of SN1987A with NuSTAR.  Since SN1987A is a young remnant, several lines from $^{194}$Os  ($\tau = 8.7$ yr) with energies around 10 keV are again prominent.  However, a detection would require large $\mathscr{O} \gtrsim 10^3$ overabundances, due chiefly to the large distance of 87A ($\approx 50$ kpc).  Prospects are better for LOFT, but detection still requires a large overabundance $\mathscr{O} \gtrsim 10^2$ (Table \ref{table:totalLOFT}).  

To summarize, the most promising currently known Galactic SN remnant is Vela Jr., for which LOFT requires an overabundance $\mathscr{O}\gtrsim 80$.  More promising is the next Galactic SN remnant, for which LOFT could detect the decay of  $^{194}$Os with near certainty for fiducial abundances.  For older remnants $^{126}$Sn ($\tau = 3.3\times 10^{5}$ yr) is the most promising isotope, despite the fact that its most important lines are at energies $\sim$ 25 keV above the peak in the LOFT/NuSTAR effective area (no long-lived nuclei have lines around 10 keV).  We were also unable to identify parent nuclei with intermediate lifetimes $10^2$ yrs $ \leq \tau \leq 10^4$ yrs with decay lines with $8$ keV $\leq \epsilon_X \leq 80$ keV. 

\subsection{Magnetar Supernova Remnants}

Figure \ref{figure:LOFTNSM} shows the required overabundance for detecting $r$-process lines from the magnetar remnant 1E 2259+586 with LOFT, normalized to the $r$-process yields predicted for magneto-centrifugal supernovae.  As in the case of Vela Jr., $^{126}$Sn ($\tau = 3.3\times 10^{5}$ years) is the most promising parent due to its long half-life, for which we find that the 3.7 keV line complex is detectable with moderately high confidence ($\sim 3 \sigma$).  A (non-)detection of such lines could thus in principle be used to test whether this magnetar was indeed born extremely rapidly spinning.

Several other of the magnetar remnants are similarly promising.  The remnant of 1E1547.0-5408 is detectable at a similar level to that of IE2259+586, while the other magnetar candidates require relatively modest overabundances $\mathscr{O} \gtrsim $few to 10 for a 2$\sigma$ detection by LOFT (Table \ref{table:totalLOFT}).  Magnetar remnants are among the most promising sources in our sample due to their combination of (1) young ages; (2) relatively close distances; (3) large assumed $r$-process yields; (4) and well-determined positions with modest angular sizes. 

 

\subsection{Neutron Star Merger Remnant}

Figure \ref{figure:LOFTNSM} shows the required overabundance for detecting a hypothetical NS merger of age $10^4$ years old and distance 10 kpc with LOFT ($\S\ref{sec:merger}$).  As in the case of Vela Jr., $^{126}$Sn ($\tau = 3.3\times 10^{5}$ years) is the most promising parent due to its long half-life, but an overabundance $\mathscr{O} \gtrsim 100$ is required for detection.   Unlike in the case of SNe, such a large overabundance can probably be ruled out because it would correspond to an actual ejecta mass $\gtrsim M_{\odot}$ that greatly exceeds the amount of mass unbound following the merger of two neutron stars (e.g.~\citealt{2013PhRvD..87b4001H}; \citealt{Fernandez&Metzger13}).  A main reason that detection is so pessimistic despite the large predicted ejecta mass in NSMs is the large background of the entire FoV of the detector, which must be overcome to search the entire Galactic plane.  


\section{Discussion and Conclusions}
\label{sec:discussion}

Building on the initial work of \citet{Qian+98}, we have investigated the prospects for detecting X-ray lines from the radioactive decay of $r$-process isotopes from supernova and neutron star binary merger remnants with current (NuSTAR) and potential future (LOFT) X-ray satellites.  Given the very low Galactic abundances of $r$-process elements, producing a detectable X-ray line signal requires a combination of fortuitous circumstances, including a decay timescale comparable to the age of the system; a large decay probability for the line of interest; and that the X-ray energy be near the peak sensitivity of the detector.  Given the very large number of $r$-process elements, it was hoped that one would possess all of these desirable properties.  Unfortunately, in most cases this turned out not to be realized.  The main obstacle to detection is that, although LOFT is planned to be significantly more sensitive that NuSTAR at energies $\sim 10$ keV near its sensitivity peak, the most promising lines from the $r$-process occur at higher energies (e.g.~$\sim 27$ keV in the case of $^{126}$Sn).  An ideal telescope to search for the $r$-process would thus combine the high effective area of LOFT with the spectral coverage and resolution extending to higher energies, as characterizes NuSTAR.      

We find that the detection of $r$-process elements in any current Galactic SNe remnant is unlikely, unless the overabundance is quite high, typically $\mathscr{O} \gtrsim 10^{3}-10^{4}$ for NuSTAR and $\mathscr{O} \gtrsim 10^{2}$ for LOFT (Tables \ref{table:totalLOFT}, \ref{table:totalNuSTAR}).  This overabundance is defined as the ratio of the required $r$-process mass in the remnant for a detection (2$\sigma$) to the expected average quantity per SN assuming that SN indeed contribute the bulk of the Galactic $r$-process.  The most promising known SN remnant is Vela Jr. (Fig.~\ref{figure:LOFTVelaJr}), for which an overabundance $\mathscr{O} \gtrsim 80$ relative to the expected average yield per SN is required for detection by LOFT.  These required overabundances may be reduced by a factor of a few due to the enhanced sensitivity achieved by combining the signal from multiple $r$-process lines (Appendix \ref{sec:multiple}; see final column of Tables \ref{table:totalLOFT}, \ref{table:totalNuSTAR}). 

Although an overabundance $\mathscr{O} \gtrsim 100$ is high compared to those expected in the normal proto-neutron star winds from core collapse SNe, much higher $r$-process yields are possible in the case of magneto-centrifugally driven explosions (e.g.~\citealt{Metzger+08}; \citealt{Ono+12}; \citealt{Winteler+12}).  It is thought that the rapidly rotating neutron stars produced by such events may be the progenitors of Galactic magnetars \citep{Thompson&Duncan93}, but there is currently no direct proof of this hypothesis.  Motivated thus, we also considered the prospects for detecting $r$-process lines from the SN remnants of Galactic magnetar, under the assumption of $r$-process yields similar to those predicted by current models of magneto-centrifugally driven SNe (\citealt{Winteler+12}).  

We find that several overlapping lines of $^{126}$Sn in the energy range $\sim 3.6-3.9$ keV are potentially detectable at 3$\sigma$ confidence for the magnetar remnants 1E1547.0-5408 and 1E2259+586 by LOFT (Fig.~\ref{figure:LOFT1E2259} and Table \ref{table:totalLOFT}). {\it Thus, the (non-)detection of such r-process lines in magnetar remnants by future X-ray telescopes could be used to (indirectly) constrain the birth periods of the magnetars.}  If such lines are detected, the millisecond birth period of the magnetar in question would be confirmed, since otherwise such high $r$-process abundances could not be achieved.  Constraining upper limits on such emission would, on the other hand, indicate that the magnetar was not created in a magneto-rotational SNe, assuming that the nucleosynthetic yields predicted by current simulations are accurate (\citealt{Winteler+12}).  Due to our simplified nature of our statistical analysis, and the neglect of possible astrophysical backgrounds, additional work is required on a system by system basis to determine what remnants are the most promising to observe and apply this test.

Mass ejected following the binary merger of two neutron stars produced similarly high $r$-process yield as magneto-rotational SNe, which motivated us to consider whether X-ray line emission could be detectable from the most recent Galactic NS merger remnant (age $t \sim 10^{4}$ yrs).  However, our lack of knowledge of the location of the most recent merger demands that such a search take the form of a systematic survey of the Galactic plane with a wide FoV telescope such as LOFT.  The much larger background signal across the whole FoV of LOFT makes detection unlikely (Figs.~\ref{figure:LOFTNSM}).  Future information on the location of the most recent merger site would substantially improve the prospects of detection by reducing this background, although it is unclear at present how such information could be obtained.  A more promising way to probe the contribution of binary NS merger to the $r$-process is via the direct detection of their decay-powered IR/optical counterparts (`kilonovae'; e.g.~\citealt{Metzger+10}), as is possible with a search triggered by the detection of a gamma-ray burst (\citealt{Berger+13}; \citealt{Tanvir+13}) or, eventually, a gravitational wave chirp.

Finally, if a supernova were to occur within our Galaxy when NuSTAR and/or LOFT is functional, we find that a $10^{6}$ second observation by either could potentially detect $r$-process lines, even assuming $r$-process yields similar to their average yield per supernova (Figs.~\ref{figure:NuSTAR3yrs10kpc}, \ref{figure:LOFT3yrs10kpc}).  This provides a relatively clean probe of whether average core collapse SNe are indeed a significant $r$-process source.  We again emphasize that our analysis does not account for intrinsic astrophysical noise, such as the luminous X-ray continuum that accompany some young SN remnants (e.g.~SN1987A).  Additional work is necessary to quantify how a realistic background might hinder extraction of the line signal over the estimates provided here.  


\begin{table}
\setlength{\tabcolsep}{3pt}
\begin{scriptsize}
\begin{center}
\vspace{-0.1 in}\caption{Detectability of $r$-process lines in SN/NSM remnants by LOFT}
\vspace{-0.3 cm}
\label{table:totalLOFT}
\begin{tabular}{ccccccc}
\hline \hline
\multicolumn{1}{c}{Source} &
\multicolumn{1}{c}{Parent$^{(a)}$} &
\multicolumn{1}{c}{$\epsilon_X$$^{(b)}$} &
\multicolumn{1}{c}{$\langle \dot{N}_{X} \rangle A_X$$^{(c)}$} &
\multicolumn{1}{c}{$\nu_X A_X$$^{(d)}$} &
\multicolumn{1}{c}{$\mathscr{O}_X$$^{(e)}$} &
\multicolumn{1}{c}{$\mathscr{O}$$^{(f)}$} \\
\hline 
&& (keV) & ($\mathrm{s}^{-1}$) & ($\mathrm{s}^{-1}$) \\
\hline 
 Vela Jr & & & & & & $27$ \\ 
	&$^{126}$Sn & $3.7$*& $1.64\times 10^{-4}$ & $16.0$ & $84$  \\ 
	&$^{126}$Sn & $4.5$*& $2.51\times 10^{-5}$ & $22.1$ & $4.9\times 10^{2}$  \\ 
	&$^{126}$Sn & $4.1$*& $2.12\times 10^{-5}$ & $24.4$ & $6.1\times 10^{2}$  \\ 
\hline 
 Cas A & & & & & & $1.5\times 10^{2}$ \\ 
	&$^{126}$Sn & $3.7$*& $2.42\times 10^{-6}$ & $0.111$ & $3.6\times 10^{2}$  \\ 
	&$^{126}$Sn & $4.5$*& $3.69\times 10^{-7}$ & $0.154$ & $2.8\times 10^{3}$  \\ 
	&$^{126}$Sn & $4.1$*& $3.12\times 10^{-7}$ & $0.170$ & $3.5\times 10^{3}$  \\ 
\hline 
 SN 1987A & & & & & & $1.4\times 10^{2}$ \\ 
	&$^{194}$Os & $10.7$*& $3.28\times 10^{-6}$ & $6.48\times 10^{-2}$ & $2.0\times 10^{2}$  \\ 
	&$^{194}$Os & $9.2$*& $2.56\times 10^{-6}$ & $7.58\times 10^{-2}$ & $2.8\times 10^{2}$  \\ 
	&$^{194}$Os & $10.9$*& $5.82\times 10^{-7}$ & $6.35\times 10^{-2}$ & $1.1\times 10^{3}$  \\ 
\hline 
 SN 1006 & & & & & & $2.1\times 10^{2}$ \\ 
	&$^{126}$Sn & $3.7$*& $1.03\times 10^{-5}$ & $4.00$ & $5.1\times 10^{2}$  \\ 
	&$^{126}$Sn & $4.5$*& $1.57\times 10^{-6}$ & $5.53$ & $3.9\times 10^{3}$  \\ 
	&$^{126}$Sn & $4.1$*& $1.32\times 10^{-6}$ & $6.10$ & $4.9\times 10^{3}$  \\ 
\hline 
 G001.9+00.3 & & & & & & $8.2\times 10^{2}$ \\ 
	&$^{126}$Sn & $3.7$*& $3.65\times 10^{-7}$ & $1.11\times 10^{-1}$ & $2.4\times 10^{3}$  \\ 
	&$^{137}$Cs & $4.5$*& $2.67\times 10^{-7}$ & $1.56\times 10^{-1}$ & $3.9\times 10^{3}$  \\ 
	&$^{137}$Cs & $4.9$*& $2.42\times 10^{-7}$ & $1.42\times 10^{-1}$ & $4.1\times 10^{3}$  \\ 
\hline 
 3yrs, 10kpc & & & & & & $0.2$ \\ 
	&$^{194}$Os & $10.7$*& $3.95\times 10^{-3}$ & $6.48\times 10^{-2}$ & $0.20$  \\ 
	&$^{194}$Os & $9.2$*& $3.08\times 10^{-3}$ & $7.58\times 10^{-2}$ & $0.20$  \\ 
	&$^{194}$Os & $10.9$*& $6.99\times 10^{-4}$ & $6.35\times 10^{-2}$ & $1.0$  \\ 
\hline 
 1E 2259+586 & & & & & & $0.50$ \\ 
	&$^{126}$Sn & $3.7$*& $8.50\times 10^{-4}$ & $0.111$ & $1.0$  \\ 
	&$^{126}$Sn & $4.5$*& $1.30\times 10^{-4}$ & $0.154$ & $7.9$  \\ 
	&$^{126}$Sn & $4.1$*& $1.10\times 10^{-4}$ & $0.170$ & $9.8$  \\ 
\hline 
 SGR 0501+4516 & & & & & & $1.5$ \\ 
	&$^{126}$Sn & $3.7$*& $3.01\times 10^{-3}$ & $16.0$ & $3.5$  \\ 
	&$^{126}$Sn & $4.5$*& $4.60\times 10^{-4}$ & $22.1$ & $27$  \\ 
	&$^{126}$Sn & $4.1$*& $3.89\times 10^{-4}$ & $24.4$ & $34$  \\ 
\hline 
 1E 1841-045 & & & & & & $2.3$ \\ 
	&$^{126}$Sn & $3.7$*& $1.61\times 10^{-4}$ & $0.111$ & $5.4$  \\ 
	&$^{126}$Sn & $4.5$*& $2.45\times 10^{-5}$ & $0.154$ & $42$  \\ 
	&$^{126}$Sn & $4.1$*& $2.07\times 10^{-5}$ & $0.170$ & $53$  \\ 
\hline 
 1E 1547.0-5408 & & & & & & $0.70$ \\ 
	&$^{126}$Sn & $3.7$*& $5.75\times 10^{-4}$ & $0.111$ & $1.5$  \\ 
	&$^{126}$Sn & $4.5$*& $8.77\times 10^{-5}$ & $0.154$ & $12$  \\ 
	&$^{126}$Sn & $4.1$*& $7.41\times 10^{-5}$ & $0.170$ & $15$  \\ 
\hline 
 AX J1845-0258 & & & & & & $4.8$ \\ 
	&$^{126}$Sn & $3.7$*& $7.75\times 10^{-5}$ & $0.111$ & $12$  \\ 
	&$^{126}$Sn & $4.5$*& $1.18\times 10^{-5}$ & $0.154$ & $87$  \\ 
	&$^{126}$Sn & $4.1$*& $9.99\times 10^{-6}$ & $0.170$ & $1.1\times 10^{2}$  \\ 
\hline 
 4U 0142+61 & & & & & & $2.4$ \\ 
	&$^{126}$Sn & $3.7$*& $6.08\times 10^{-4}$ & $1.78$ & $5.7$  \\ 
	&$^{126}$Sn & $4.5$*& $9.28\times 10^{-5}$ & $2.46$ & $45$  \\ 
	&$^{126}$Sn & $4.1$*& $7.85\times 10^{-5}$ & $2.71$ & $55$  \\ 
\hline 
 CXOU J171405.7-381031 & & & & & & $2.7$ \\ 
	&$^{126}$Sn & $3.7$*& $1.61\times 10^{-4}$ & $0.160$ & $6.5$  \\ 
	&$^{126}$Sn & $4.5$*& $2.45\times 10^{-5}$ & $0.221$ & $51$  \\ 
	&$^{126}$Sn & $4.1$*& $2.07\times 10^{-5}$ & $0.244$ & $63$  \\ 
\hline 
 SGR 0526-66 & & & & & & $1.2\times 10^{2}$ \\ 
	&$^{126}$Sn & $3.7$*& $3.08\times 10^{-6}$ & $0.111$ & $2.8\times 10^{2}$  \\ 
	&$^{126}$Sn & $4.5$*& $4.70\times 10^{-7}$ & $0.154$ & $2.2\times 10^{3}$  \\ 
	&$^{126}$Sn & $4.1$*& $3.97\times 10^{-7}$ & $0.170$ & $2.7\times 10^{3}$  \\ 
\hline 
 NS Merger & & & & & & $55$ \\ 
	&$^{126}$Sn & $3.7$*& $8.42\times 10^{-5}$ & $16.0$ & $1.2\times 10^{2}$  \\ 
	&$^{126}$Sn & $4.5$*& $1.28\times 10^{-5}$ & $22.1$ & $9.6\times 10^{2}$  \\ 
	&$^{126}$Sn & $4.1$*& $1.09\times 10^{-5}$ & $24.4$ & $1.2\times 10^{3}$  \\ 
\hline
\hline
\end{tabular}
\end{center}
$^{(a)}$ see Table \ref{table:parents} for the daughter isotope. $^{(b)}$decay line energy. $^{(c)}$fiducial flux of decay line photons (eq.~\ref{equation:flux}) multiplied by the detector effective area at the energy $\epsilon_X$. $^{(d)}$detector noise multiplied by the detector effective area at energy $\epsilon_X$. $^{(e)}$Overabundance needed to achieve $\mathscr{P}=0.95$ detection confidence for specific line. $^{(f)}$Overabundance  needed to achieve $\mathscr{P}=0.95$ detection confidence by combining all lines (Appendix A). *average of overlapping lines within spectral resolution of LOFT.$^{(g)}$ 
\end{scriptsize}
\end{table}

\begin{table}
\setlength{\tabcolsep}{3pt}
\begin{scriptsize}
\begin{center}
\vspace{-0.1 in}\caption{Detectability of $r$-process lines in SN/NSM remnants by NuSTAR}
\vspace{-0.3cm}
\label{table:totalNuSTAR}
\begin{tabular}{cccccccc}
\hline \hline
\multicolumn{1}{c}{Source} &
\multicolumn{1}{c}{Parent$^{(a)}$} &
\multicolumn{1}{c}{$\epsilon_X$$^{(b)}$} &
\multicolumn{1}{c}{$\langle \dot{N}_{X} \rangle A_X$$^{(c)}$} &
\multicolumn{1}{c}{$\nu_X A_X$$^{(d)}$} &
\multicolumn{1}{c}{$\mathscr{O}_X$$^{(e)}$} &
\multicolumn{1}{c}{$\mathscr{O}$$^{(f)}$} \\
\hline 
& & (keV) & ($\mathrm{s}^{-1}$) & ($\mathrm{s}^{-1}$) \\
\hline 
 Vela Jr & & & & & & $5.4\times 10^{3}$ \\ 
	&$^{126}$Sn & $26.3$*& $6.40\times 10^{-8}$ & $0.150$ & $1.6\times 10^{4}$  \\ 
	&$^{126}$Sn & $64.3$ & $8.47\times 10^{-9}$ & $3.00\times 10^{-2}$ & $5.4\times 10^{4}$  \\ 
	&$^{126}$Sn & $23.3$*& $1.72\times 10^{-8}$ & $0.150$ & $5.9\times 10^{4}$  \\ 
\hline 
 Cas A & & & & & & $5.1\times 10^{3}$ \\ 
	&$^{126}$Sn & $26.3$*& $3.39\times 10^{-8}$ & $3.75\times 10^{-2}$ & $1.5\times 10^{4}$  \\ 
	&$^{126}$Sn & $64.3$ & $4.49\times 10^{-9}$ & $7.50\times 10^{-3}$ & $5.1\times 10^{4}$  \\ 
	&$^{126}$Sn & $23.3$*& $9.11\times 10^{-9}$ & $3.75\times 10^{-2}$ & $5.6\times 10^{4}$  \\ 
\hline 
 SN 1987A & & & & & & $1.7\times 10^{3}$ \\ 
	&$^{194}$Os & $9.2$*& $2.59\times 10^{-8}$ & $1.00\times 10^{-3}$ & $3.3\times 10^{3}$  \\ 
	&$^{194}$Os & $10.7$*& $1.16\times 10^{-8}$ & $3.00\times 10^{-4}$ & $4.1\times 10^{3}$  \\ 
	&$^{194}$Cs & $12.8$*& $6.65\times 10^{-9}$ & $2.50\times 10^{-4}$ & $6.5\times 10^{3}$  \\ 
\hline 
 SN 1006 & & & & & & $2.2\times 10^{4}$ \\ 
	&$^{126}$Sn & $26.3$*& $1.60\times 10^{-8}$ & $0.150$ & $6.4\times 10^{4}$  \\ 
	&$^{126}$Sn & $64.3$ & $2.11\times 10^{-9}$ & $3.00\times 10^{-2}$ & $2.2\times 10^{5}$  \\ 
	&$^{126}$Sn & $23.3$*& $4.29\times 10^{-9}$ & $0.150$ & $2.4\times 10^{5}$  \\ 
\hline 
 G001.9+00.3 & & & & & & $8.3\times 10^{3}$ \\ 
	&$^{126}$Sn & $26.3$*& $5.12\times 10^{-9}$ & $3.37\times 10^{-3}$ & $3.0\times 10^{4}$  \\ 
	&$^{137}$Cs & $32.1$*& $3.87\times 10^{-9}$ & $2.25\times 10^{-3}$ & $3.2\times 10^{4}$  \\ 
	&$^{137}$Cs & $36.4$*& $6.89\times 10^{-10}$ & $6.75\times 10^{-4}$ & $1.0\times 10^{5}$  \\ 
\hline 
 3yrs, 10kpc & & & & & & $0.90$ \\ 
	&$^{125}$Sb & $27.4$*& $3.14\times 10^{-5}$ & $3.75\times 10^{-4}$ & $1.7$  \\ 
	&$^{125}$Sb & $19.9$*& $3.12\times 10^{-5}$ & $1.00\times 10^{-3}$ & $2.7$  \\ 
	&$^{194}$Os & $9.2$*& $1.39\times 10^{-5}$ & $3.00\times 10^{-4}$ & $4.4$  \\ 
\hline 
 1E 2259+586 & & & & & & $5.8$ \\ 
	&$^{126}$Sn & $26.3$*& $1.19\times 10^{-5}$ & $5.42\times 10^{-3}$ & $20$  \\ 
	&$^{126}$Sn & $23.3$*& $3.20\times 10^{-6}$ & $5.42\times 10^{-3}$ & $70$  \\ 
	&$^{126}$Sn & $29.8$*& $2.41\times 10^{-6}$ & $1.08\times 10^{-2}$ & $1.2\times 10^{2}$  \\ 
\hline 
 SGR 0501+4516 & & & & & & $3.0\times 10^{2}$ \\ 
	&$^{126}$Sn & $26.3$*& $1.17\times 10^{-6}$ & $0.150$ & $8.7\times 10^{2}$  \\ 
	&$^{126}$Sn & $23.3$*& $3.15\times 10^{-7}$ & $0.150$ & $3.3\times 10^{3}$  \\ 
	&$^{126}$Sn & $64.3$ & $1.55\times 10^{-7}$ & $3.00\times 10^{-2}$ & $3.0\times 10^{3}$  \\ 
\hline 
 1E 1841-045 & & & & & & $33$ \\ 
	&$^{126}$Sn & $26.3$*& $2.25\times 10^{-6}$ & $6.62\times 10^{-3}$ & $1.0\times 10^{2}$  \\ 
	&$^{126}$Sn & $23.3$*& $6.05\times 10^{-7}$ & $6.62\times 10^{-3}$ & $3.6\times 10^{2}$  \\ 
	&$^{126}$Sn & $29.8$*& $4.56\times 10^{-7}$ & $1.32\times 10^{-2}$ & $6.7\times 10^{2}$  \\ 
\hline 
 1E 1547.0-5408 & & & & & & $22$ \\ 
	&$^{126}$Sn & $26.3$*& $8.06\times 10^{-6}$ & $3.75\times 10^{-2}$ & $70$  \\ 
	&$^{126}$Sn & $23.3$*& $2.16\times 10^{-6}$ & $3.75\times 10^{-2}$ & $2.4\times 10^{2}$  \\ 
	&$^{126}$Sn & $29.8$*& $1.63\times 10^{-6}$ & $7.50\times 10^{-2}$ & $4.4\times 10^{2}$  \\ 
\hline 
 AX J1845-0258 & & & & & & $38$ \\ 
	&$^{126}$Sn & $26.3$*& $1.09\times 10^{-6}$ & $1.95\times 10^{-3}$ & $1.1\times 10^{2}$  \\ 
	&$^{126}$Sn & $23.3$*& $2.92\times 10^{-7}$ & $1.95\times 10^{-3}$ & $4.1\times 10^{2}$  \\ 
	&$^{126}$Sn & $29.8$*& $2.20\times 10^{-7}$ & $3.90\times 10^{-3}$ & $7.6\times 10^{2}$  \\ 
\hline 
 4U 0142+61 & & & & & & $1.6\times 10^{2}$ \\ 
	&$^{126}$Sn & $26.3$*& $2.13\times 10^{-6}$ & $0.150$ & $4.8\times 10^{2}$  \\ 
	&$^{126}$Sn & $23.3$*& $5.73\times 10^{-7}$ & $0.150$ & $1.8\times 10^{3}$  \\ 
	&$^{126}$Sn & $64.3$ & $2.82\times 10^{-7}$ & $3.00\times 10^{-2}$ & $1.7\times 10^{3}$  \\ 
\hline 
 CXOU J171405.7-381031 & & & & & & $93$ \\ 
	&$^{126}$Sn & $26.3$*& $2.26\times 10^{-6}$ & $5.40\times 10^{-2}$ & $2.8\times 10^{2}$  \\ 
	&$^{126}$Sn & $23.3$*& $6.06\times 10^{-7}$ & $5.40\times 10^{-2}$ & $1.1\times 10^{3}$  \\ 
	&$^{126}$Sn & $29.8$*& $4.57\times 10^{-7}$ & $0.108$ & $1.9\times 10^{3}$  \\ 
\hline 
 SGR 0526-66 & & & & & & $4.3\times 10^{2}$ \\ 
	&$^{126}$Sn & $26.3$*& $4.32\times 10^{-8}$ & $3.75\times 10^{-4}$ & $1.3\times 10^{3}$  \\ 
	&$^{126}$Sn & $23.3$*& $1.16\times 10^{-8}$ & $3.75\times 10^{-4}$ & $4.6\times 10^{3}$  \\ 
	&$^{126}$Sn & $29.8$*& $8.75\times 10^{-9}$ & $7.50\times 10^{-4}$ & $8.4\times 10^{3}$  \\ 
\hline 
 NS Merger & & & & & & $1.1\times 10^{4}$ \\ 
	&$^{126}$Sn & $26.3$*& $3.28\times 10^{-8}$ & $1.50\times 10^{-1}$ & $3.1\times 10^{4}$  \\ 
	&$^{126}$Sn & $64.3$ & $4.34\times 10^{-9}$ & $3.00\times 10^{-2}$ & $1.0\times 10^{5}$  \\ 
	&$^{126}$Sn & $23.3$*& $8.81\times 10^{-9}$ & $1.50\times 10^{-1}$ & $1.2\times 10^{5}$  \\ 
\hline
\hline
\end{tabular}
\end{center}
Same as Table \ref{table:totalLOFT}, but for NuSTAR instead of LOFT.
\end{scriptsize}
\end{table}
\newpage

Is it reasonable to expect a Galactic SNe during the lifetime of NuSTAR or the next generation of X-ray telescopes such as LOFT?  The galactic SNe rate range up to 10 per century, and there is a 50\% probability that the next SN will occur within 10 kpc of Earth (\citealt{Adams:2013ana}). Modeling the occurrence of SNe as a Poisson process gives the approximate probability of at least one new SN occurring in the next $t$ years within 10 kpc as:
\begin{equation}
\label{equation:poissonccne}
P(n \geq 1) \simeq 1 - e^{-t/20}
\end{equation}
Since NuSTAR may be operational as late as 2022 (\citealt{Harrison+13}) the probability of a nearby core collapse event during its operational lifetime is $\sim$26\% (assuming a 3 year window for the SN ejecta to become transparent). Assuming LOFT has a similar lifetime ($\sim$ 10 years) as NuSTAR, the probability that LOFT will be operational during the next SN is $\sim$30\%. 

In conclusion, X-ray decay lines represent a potentially powerful probe of the origin of $r$-process nuclei and the birth period of magnetars.  This scientific opportunity also provides an additional motivation for the next generation of X-ray satellites.  The very low expected fluxes will require, however, carefully planned observations to enhance signal and minimize background, and a well-defined statistical criteria for detection, based on the expected yield of $r$-process nuclei and their decay channels.  The results of this paper provide a framework and motivation for the additional work that will be necessary to perform such a test using data from actual remnants.

\section*{Acknowledgments}

JR acknowledges support through the Columbia Undergraduate Scholars Program.  JR and BDM acknowledge support from the Department of Physics at Columbia University.  A.A. is supported by the Helmholtz-University Young Investigator grant No. VH-NG-825.  GMP is partly supported by the Deutsche Forschungsgemeinschaft through contract SFB 634, the Helmholtz International Center for FAIR within the framework of the LOEWE program launched by the state of Hesse, and the Helmholtz Association through the Nuclear Astrophysical Virtual Institute (VH-VI-417).

\appendix

\section{T-Statistic For Assessing Line Sensitivity}

\label{sec:append}

A rough approximation for the signal to noise level required for a statistically significant detection of an $r$-process X-ray decay line can be developed using Student's t-test (\citealt{james:06}):
\begin{equation}
 \label{t-test}
	t = \frac{\bar{X} - \mu}{\sqrt{\frac{\sigma_{\rm X}^2}{N_{\rm X}} + \frac{\sigma_{\mu}^2}{N_{\mu}}}} 
\end{equation}
where $\bar{X}$ and $\sigma_{X}$ are the measured mean and standard deviation, and $N_{\rm X}$ is the number of trials of the measurement performed. Here $\mu$ and $\sigma_{\mu}$ are the expected (null hypothesis) means and standard deviation for $N_{\mu}$ trials. 

We apply the t-statistic to the detection of X-ray decay photons by making the following associations
\begin{equation} \label{equation:poisson}
\begin{aligned}
	&  N = \frac{T}{\Delta t} \\
	& \bar{X} = \sigma_{\rm X}^2 = (\dot{N}_{X} + \nu_X)\Delta t A_{X} \\
	& \mu = \sigma_{\mu}^2 = \nu_X \Delta t A_{X},
       \end{aligned}
\end{equation}
where $\Delta t$ is the time resolution of the detector, $N = T/{\Delta t}$ is the number of data points collected ("degrees of freedom") over the total time $T$ of the observations, $\mu_X$ is the average background rate per unit time per unit area in the frequency bin $X$; and $\dot{N}_X$ is the rate of signal photons (eq.~[\ref{equation:flux}]).

As the distribution of signal and noise photons are assumed to follow a Poisson distribution, the probability of the detector registering two photons simultaneously at the energy $X$ and effective area $A_{X}$ is 
\begin{equation} \label{equation:over2}
	P(n \geq 2) = A_{X}(1 - (1+\lambda)e^{-\lambda})
\end{equation}
where $\lambda = (\dot{N}_{X}+\nu_X)\Delta t$. Typical values for the variables in equation \ref{equation:over2} are $A_{X} \approx 10^4~ \mathrm{cm}^2$, $\dot{N}_{X}+\nu_X \leq 1 \mathrm{cts/ sec~cm}^2$, and $\Delta t = 10 \mu$s, so an upper limit of the probability of a double photon count in any measurement during one time bin is approximately $\sim 10^{-7}$. Thus, the assumption that of only single photon is measured during each time interval is valid. The t-test (equation \ref{t-test}) then reduces to:
\begin{equation} \label{equation:t-test-exp}
	t_{X} = \frac{\dot{N}_{X} \sqrt{A_{X}T}}{\sqrt{\dot{N}_{X} + 2 \nu_X}}
\end{equation}
For $N \rightarrow \infty$ ($\Delta t \ll T$), a t value exceeding 1.9 corresponds to a $2 \sigma$ confidence that the measurement can be distinguished from the null, which in this case is pure noise.



\subsection{Increasing detection sensitivity by combining the signal of multiple $r$-process lines}
\label{sec:multiple}

Even if the probability of detecting individual X-ray lines from $r$-process elements for a given source is very small, a greater sensitivity could be achieved by simultaneously searching for multiple lines, either from the same isotope or from different isotopes produced in coincidence.  

Let the confidence of detecting an {\it anomalous} signal at energy $\epsilon_{X_n}$ with flux $\dot{N}_{X_n}$ above the noise be $\mathscr{C}_{X_n}$ (this confidence is determined from its t-statistic $t_{X_n}$ calculated according to eq.~[\ref{equation:ttestexp}]).  If no $r$-process parent nuclei in fact produced such decay lines with X-ray energies $\epsilon_{X_1},...,\epsilon_{X_N}$, then the probability of detecting all of these lines, each with a flux greater than the mean background noise $\nu_{X_n}$, is given by
\begin{equation}
\label{equation:tc}
\mathscr{C} =\prod_{n=1}^N \mathscr{C}_{X_n}
\end{equation} 
As each $\mathscr{C}_{X_n} \leq 1$,  the confidence for the random detection of an $r$-process site being falsely discovered goes down as the number of possible detection lines increases. The probability $\mathscr{P}$ that the total number of lines not being due to random fluctuations is then:
\begin{equation}
\label{equation:tp}
\mathscr{P} = (1- \mathscr{C}) = \left(1- \prod_{n=1}^N \mathscr{C}_{X_n} \right)
\end{equation} 
This total probability $\mathscr{P}$ thus quantifies the likelihood that $r$-process nucleosynthesis (considered as a whole) occurred in the astrophysical site under observation.  The required over abundance for a 2$\sigma$ confidence detection $\mathscr{O}$ is shown in the final column Tables \ref{table:totalLOFT} and \ref{table:totalNuSTAR}, which can be lower by up to an order of magnitude as compared to that required based on the best individual X-ray lines.  



\end{document}